\newcommand{\tchi}{\tilde{\chi}}

\documentstyle[epic,eepic,12pt]{article}
\renewcommand{\theequation}{\thesection.\arabic{equation}}
\newcounter{saveeqn}
\newcommand{\alpheqn}{\setcounter{saveeqn}{\value{equation}}%
\stepcounter{saveeqn}\setcounter{equation}{0}%
\renewcommand{\theequation}{\mbox{\thesection.\arabic{saveeqn}-\alph{equation}}}}
\newcommand{\reseteqn}{\setcounter{equation}{\value{saveeqn}}%
\renewcommand{\theequation}{\thesection.\arabic{equation}}}
\begin{document}
\bibliographystyle{unsrt}
\title{Dynamical Fluctuating Charge Force Fields: Application to
Liquid Water}
\author{Steven W. Rick, Steven J. Stuart, and B.J. Berne \\
Department of Chemistry and Center for Biomolecular Simulation\\
Columbia University, NY 10027}

\date{\ }

\maketitle

\begin{abstract}
A new molecular dynamics model
in which the point charges on atomic sites are
allowed to fluctuate in response to the environment is developed and
applied to water.  The idea for treating charges as variables is based
on the concept of electronegativity equalization according to which:
(a) The electronegativity of an atomic site is dependent on the atom's
type and charge and is perturbed by the electrostatic potential it experiences
from its neighbors and (b) Charge is transferred between atomic sites
in such a way that electronegativities are equalized.  The charges are
treated as dynamical variables
 using an extended Lagrangian method in which the charges
are given a fictitious mass, velocities and kinetic energy and then
propagated according to Newtonian mechanics along with the atomic
degrees of freedom.  Models for water with fluctuating charges are
developed using the geometries of two common fixed-charge water
potentials: the simple point charge (SPC) and the
4-point transferable intermolecular
potential (TIP4P). Both fluctuating charge models give
accurate predictions for gas-phase and
liquid state properties, including radial
distribution functions, the dielectric constant, and the diffusion
constant.  The method does not introduce any new intermolecular
interactions beyond those already present in the fixed charge models
and increases the computer time by only a factor of 1.1, making this
method tractable for large systems.
\end{abstract}

\vspace{.5in}

\newpage

\baselineskip=18pt plus 3pt
\lineskip=3pt minus 1 pt
\lineskiplimit=2pt

\section{Introduction}

In simple molecular force fields the intramolecular electronic
structure is often modeled by point charges fixed on well defined
sites in the molecular frame. The charges are constant and thus cannot
change in response to changing electrostatic fields which arise from
movement of the atoms during the simulation. In reality, molecular
electronic structure can be strongly influenced by the molecular
environment. For example the total dipole moment of water changes from
1.85 Debye in the gas phase to approximately 2.5 Debye in the liquid
phase. Thus the charges used in simulations based on fixed charge
force fields must reflect average or mean field charge values for the
particular phase and are not transferable to different
thermodynamic states or to different media. In addition, the
self-energy involved in the change in charge accompanying the
transition from gas phase to liquid phase is commonly neglected in
most force field parametrizations. This ``missing term'' in fixed
charge pair potentials can be significant (2 to 5 kcal/mole for
water)\cite{Berendsen}. Charge induction effects are not pair-wise
additive and improved models must go beyond pair potentials. The
purpose of this paper is to present a new simulation method in which
the charges are responsive to environmental changes.

The approach taken here combines the electronegativity equalization
(EE) method for solving for atomic charges and the extended Lagrangian
method for treating fictitious degrees of freedom as dynamical
variables.  The calculation of atomic electronegativities using
density functional theory, basis set methods, or empirical data and
the use of this information to estimate charges for large molecules
has a long history\cite{ParrYang,Allinger,Mortier,No,Rappe,Richards}.
The electronegativity of an atomic
site is dependent on its charge and the electronegativities of the
neighboring atoms.  Parr has shown that the Mulliken electronegativity
($\chi_i$) of an isolated atom $i$ is the negative of the chemical
potential ($\mu_i$) of the electron gas surrounding its nucleus,
\begin{equation}
\mu_i=\frac{\partial
E}{\partial N} =-\chi_i = -e\frac{\partial E}{\partial Q_i}
\label{eq:parr}
\end{equation}
where E is the ground state energy, $N$ is the number of electrons in
the atom (treated as a continuous variable), $Q$ is the charge on
the atom, and $e$ is the elementary charge.
Use has been made of the fact that $Q$ is related to $N$
by $Q=-e(N-Z)$ where $Z$
is the atomic number of the atom. In a many atom system, the electron
gas will equilibrate with the instantaneous positions of the nuclei in
such a way that the electrochemical potential of the electron gas
will be equal at all atomic sites.
In this picture,
electrons will then move among atoms from regions of low
electronegativity (or high electrochemical potential) to regions of
high electronegativity (low electrochemical potential). For the ground
state electronic configuration, the electrochemical potentials are
equal. The approach we take is to treat charges on the molecular sites
as dynamical variables by introducing fictitious kinetic energy terms
and self energy terms for these charges into the Lagrangian for the
system along with Lagrange constraints representing various conditions
of electroneutrality. In this extended Lagrangian
approach\cite{Andersen,ParrinelloRahman,Car,Nose} the charges are
propagated according to Newtonian mechanics in a similar way to the
atomic degrees of freedom.
Although the fluctuating charge model
provides a convenient starting point for the discussion of complex
solutes like n-methylacetamide and proteins, we focus
here on its application to the simulation of neat water.

Liquid water was chosen for the first application of this fluctuating
charge (or fluc-q) model because charge polarization effects should be
important for water. Simple water models, such as the simple point charge
(SPC)\cite{BerendsenSPC} or the 4-point transferable intermolecular potential
TIP4P\cite{JorgensenKlein}, with a
Lennard-Jones interaction between oxygen atoms and three charge sites
with fixed liquid state charges, can give accurate predictions for
many equilibrium properties of liquid water, including the energy and
radial distribution functions. Due to their simplicity and relative
accuracy, these are perhaps the two most widely used water
potentials. However, the translational and rotational time scales of
these models are too fast, although
the SPC/E\cite{Berendsen}
reparameterization gives improved relaxation times.
More importantly, these models are
unable to deal accurately with heterogeneous environments.
Simulations
of sodium octanoate micelles in SPC water predict too much penetration
of water molecules into the micelle\cite{Jonsson,Ferrario} in
contrast to experiments\cite{Dill}. The purpose of this paper is to
add fluctuating charges to the SPC and TIP4P potentials and to thus
devise a model which has improved static and dynamical properties
relative to the fixed charge models and which should is easily
extendible to more complex solutions.

Electrical induction can also be described to lowest order using
fixed gas phase charges and point polarizabilities. Many
dipole polarization models have been used to simulate liquid water
\cite{StillingerDavid,Barnes,Rullman,SprikKlein,Ahlstrom,Cieplak,WallqvistBerne,Dang,BernardoLevy}.
The fluc-q model
presented here is an alternative which differs from the dipole polarizable
models in two respects. First, the fluc-q models have polarizabilities to
all orders in the charge moments and not only dipolar polarizability. In
addition, the dipole polarizability models introduce a new interaction (the
1/r$^3$ dipole-dipole interaction) and in order to solve for the induced
dipole moments, either the induced dipole equations are solved
iteratively, by matrix inversion, or the polarizations are treated in an
extended Lagrangian framework.
The iterative solution method increases the cost by a
factor of two\cite{WallqvistBerne} and the extended Lagrangian methods by a
factor of two\cite{VanBelle3} to four\cite{SprikKlein}. The fluc-q method
introduces no new intermolecular
interactions beyond the fixed-charge models and increases the cpu
time by a factor of only 1.1.
Of course the dipole
polarizability model can also be cast in terms of Drude dispersion
oscillators, a system that also lends itself to treatment by the
extended Lagrangian method.

This paper is organized as follows as follows. Section~\ref{sec:methodology}
describes the fluc-q method and the form of the water-water interactions.
Section~\ref{sec:numerical} describes our implementation
of molecular dynamics.
Section~\ref{sec:results} describes the results of these models and
Section~\ref{sec:conclusions} summarizes the conclusions.

\section{Dynamical Fluctuating Charge Models}
\setcounter{equation}{0}
\label{sec:methodology}

The central idea for treating charges as dynamical variables is based on
the electronegativities of atomic sites. Parr has shown, using Kohn-Sham
theory, that in an atom, the atomic electrons, regarded as an electron gas,
have a chemical potential which is the negative of the Mulliken
electronegativity\cite{ParrYang}. In a many-atom system the full electron
gas will distribute itself so that its electrochemical potential takes the
same value at every nuclear site. This principle of electronegativity
equalization (EE) was first proposed by Sanderson\cite{Sanderson}. If a
given site moves so that it feels a different electrostatic potential it
will take on a different charge. In this way the charges on molecular sites
will respond to the environment.

In the isolated atom the energy of creating a partial charge, Q$_A$,
can be expanded to second order as
\begin{equation}
{\rm E}({\rm Q}_A) = {\rm E}_A(0) +\tilde{\chi}_A^0 {\rm Q}_A + {1
\over 2} {\rm J}_{AA}^0 {\rm Q}_A^2
\label {eq:expan}
\end{equation}
where $\tilde{\chi}_A^0$ and ${\rm J}_{AA}^0$, are parameters
dependent on the atom type.  Values of $\tilde{\chi}_A^0$ and
J$_{AA}^0$ can be calculated using basis set, density functional
theory methods or empirical data: $\tilde{\chi}_A^0$ is the Mulliken
electronegativity (per electronic charge $e$) and J$_{AA}^0$ is twice
the hardness of the electronegativity of the isolated atom.  The
energy of a system of N$_{\rm molec}$ molecules each with N$_{\rm atom}$
atoms is
\begin{equation}
{\rm U}(\{ {\bf Q}\},\{ {\bf r} \})
= \sum_{i=1}^{N_{molec}}\sum_{\alpha=1}^{N_{atom}}
\left( {\rm E}_{\alpha}(0)
+ \tilde{\chi}_{\alpha}^0 {\rm Q}_{i\alpha} + {1 \over 2} {\rm J}_{\alpha
\alpha}^0 {\rm Q}_{i\alpha}^2 \right) + \sum_{i\alpha<j\beta} {\rm
J}_{\alpha \beta}({\rm r}_{i\alpha j\beta}) {\rm Q}_{i\alpha} {\rm Q}_{j\beta}
+ \sum_{i\alpha<j\beta}
{\rm V}({\rm r}_{i\alpha j\beta} )
\label {eq:ec}
\end{equation}
where E$_{\alpha}$(0) is the ground state energy of atom $\alpha$,
r$_{i\alpha j\beta}$ is the distance,
${\rm J}_{\alpha\beta}$(r$_{i\alpha j\beta}$) the Coulomb interaction
and V(r$_{i\alpha j\beta}$) is any additional non-Coulombic interaction
between $i\alpha$ and $j\beta$.
The electronegativity  per unit charge of atom A is given by
\begin{equation}
\tilde{\chi}_A=
\left({\partial {\rm U} \over \partial {\rm Q}_A}\right).
\label {eq:chi}
\end{equation}
The charges, by the EE principle, are then those for which the
electronegativities are equal.
This is equivalent to minimizing the energy, subject to
a charge neutrality constraint.
Since the potential is quadratic in the charges,
the minimization will lead to a set of coupled linear equations
for the charge.

The charges are not independent variables since there is a charge
conservation constraint. For uncharged
molecular systems, the constraint can be of two
types:
\alpheqn
\begin{enumerate}
\item The entire system is constrained to be
neutral, so individual molecules can carry a non-zero
charge because there can be intermolecular charge
transfer,
\begin{equation}
\sum_{i=1}^{N_{molec}} \sum_{\alpha=1}^{N_{atom}} {\rm
Q}_{i\alpha}=0.
\label{eq:totalneut}
\end{equation}
\item Each molecule is constrained to be neutral, so there is no
intermolecular charge transfer,
\begin{equation}
\sum_{\alpha=1}^{N_{atom}} {\rm
Q}_{i\alpha}=0.
\label{eq:molecneut}
\end{equation}
\end{enumerate}
\reseteqn
 With intermolecular charge
transfer, the chemical potentials of all the atoms of the system will be
equal. Without charge transfer, the chemical potential of an atom will only
be equal to the chemical potential of atoms on the same molecule. The
simplest way to treat the charge neutrality constraint is to treat the
charges as independent and use the method of undetermined multipliers to
enforce the constraint. The Lagrangians for cases 1 and 2 are:
\alpheqn
\begin{equation}
L_1=\sum_{i=1}^{N_{molec}}
\sum_{\alpha=1}^{N_{atom}} {1 \over 2} m_{\alpha} \dot{{\rm r}}_{i\alpha}^2
+ \sum_{i=1}^{N_{molec}} \sum_{\alpha=1}^{N_{atom}} {1 \over 2} {\rm
M}_{\rm Q} \dot{{\rm Q}}_{i\alpha}^2 - {\rm U}(\{{\bf Q}\},\{{\bf
r}\})-\lambda \sum_{i=1}^{N_{molec}} \sum_{\alpha=1}^{N_{atom}} {\rm
Q}_{i\alpha}
\label{eq:lagrangian1}
\end{equation}
and
\begin{equation}
L_2=\sum_{i=1}^{N_{molec}}
\sum_{\alpha=1}^{N_{atom}} {1 \over 2} m_{\alpha} \dot{{\rm r}}_{i\alpha}^2
+ \sum_{i=1}^{N_{molec}} \sum_{\alpha=1}^{N_{atom}} {1 \over 2} {\rm
M}_{\rm Q} \dot{{\rm Q}}_{i\alpha}^2 - {\rm U}(\{{\bf Q}\},\{{\bf
r}\})-\sum_{i=1}^{N_{molec}} \lambda_i \sum_{\alpha=1}^{N_{atom}} {\rm
Q}_{i\alpha}
\label{eq:lagrangian2}
\end{equation}
\reseteqn
where m$_{\alpha}$ is the mass of atom $\alpha$
and M$_{\rm Q}$ is a fictitious charge
``mass'', which has units of energy-sec$^2$/charge$^2$
and the $\lambda$'s are Lagrange multipliers.
The nuclear degrees of freedom evolve according to Newton's equation
\begin{equation}
m_{\alpha} \ddot{{\rm r}}_{i\alpha}
= - {\partial {\rm U}(\{{\bf Q}\},\{{\bf r}\})
\over \partial {\rm r}_{i\alpha} }
\label {eq:newton}
\end{equation}
and the set of charges  evolve in time according to
\begin{equation}
{\rm M}_{\rm Q}  {\rm \ddot{Q}}_{i\alpha}
= - {\partial {\rm U}(\{{\bf Q}\},\{{\bf r}\})
\over \partial {\rm Q}_{i\alpha}}
-\lambda_{i}
= -\tchi_{i\alpha} - \lambda_{i}
\label {eq:cp}
\end{equation}
where $\lambda_{i}$ is the Lagrange multiplier for
the charge neutrality constraint, given either by
Eq.~(\ref{eq:totalneut}) or Eq.~(\ref{eq:molecneut}).
It should be noted that if the total charge in the liquid
is a constant of the motion,
then
\alpheqn
\begin{equation}
\sum_{i=1}^{N_{molec}}\sum_{\alpha=1}^{N_{atom}}{\rm
\ddot{Q}}_{i\alpha}=0
\end{equation}
whereas if the total charge on molecule $i$ is a constant of the
motion, then for all $i$
\begin{equation}
\sum_{\alpha=1}^{N_{atom}}{\rm \ddot{Q}}_{i\alpha}=0.
\end{equation}
\reseteqn
Substitution of Eq.~(\ref{eq:cp}) into
each of the above two equations
yields, respectively,
\alpheqn
\begin{equation}
\lambda = - {1 \over N_{molec} N_{atom} }
\sum_{i=1}^{N_{molec}} \sum_{\alpha=1}^{N_{atom}} \tchi_{i\alpha}
\label{eq:lambda1}
\end{equation}
where $\lambda$ is equal to the negative of the average of the system's
total electronegativity, and
\begin{equation}
\lambda_i = - {1 \over N_{atom} }
\sum_{\alpha=1}^{N_{atom}} \tchi_{i\alpha}
\label{eq:lambda2}
\end{equation}
\reseteqn
where $\lambda_i$ is the negative of the average electronegativity on
molecule $i$.  Substitution of Eqs.~(\ref{eq:lambda1}) and
(\ref{eq:lambda2}) into the equations of the motion for the charges
gives:
\alpheqn
\begin{equation}
{\rm M}_{\rm Q} {\rm \ddot{Q}}_{i\alpha} =-{1 \over N_{molec} N_{atom} }
\sum_{i=1}^{N_{molec}}\sum_{\beta=1}^{N_{atom}}
\left(\tchi_{i\alpha}-\tchi_{i\beta}\right)
\end{equation}
and
\begin{equation}
{\rm M}_{\rm Q} {\rm \ddot{Q}}_{i\alpha} =- {1 \over N_{atom} }
\sum_{\beta=1}^{N_{atom}}\left(\tchi_{i\alpha}-\tchi_{i\beta}\right).
\end{equation}
\reseteqn
Whether or not charges are allowed to transfer between molecules or
just between atoms on the same molecule makes little difference in the
algorithm.  The force on the charge is simply the difference
between
the average electronegativity and
the electronegativity at that site.
For example, if the electronegativity is greater
than the average, then the force acts to decease the charge until the
electronegativities are all equal.  In the present application, we
have included a charge neutrality constraint on each water molecule,
there is no charge transfer between molecules (case 2).

There have been many different applications of the EE principle, with
differing values for the parameters $\tchi_A$ and J$_{AA}$ (some with
higher order terms) which have been applied to a variety of
molecules\cite{Allinger,Mortier,No,Rappe,Richards}. Some of these
applications have been used as potential input into simulations. However,
in order to fully treat the charge fluctuations which arise in the course
of a simulation, each new configuration would require the calculation of a new
set of charges. This can be done simply using the extended Lagrangian
method to derive the  equations of motion for Hamiltonians which depend on
auxiliary degrees of freedom\cite{Andersen,Car}.

The above approach makes no allowance for the possibility that there can be
barriers to the charge transfer preventing charge equilibration. Thus for
example two atomic sites separated by a great distance will in reality not
transfer charges because the probability of tunneling through a wide
barrier is small, yet the above model will allow charge transfer. This
poses a major problem in principle when molecules are separated in vacuo,
a problem that also exists in the Car-Parrinello method\cite{Car}.
For this reason, we have restricted the charge equilibration to
intramolecular charge transfer. It would be useful to generalize this model
to include such kinetic restrictions on charge transfer.

The charge mass, M$_{\rm Q}$, a fictitious quantity, should be chosen to be
small
enough to guarantee that the charges readjust very rapidly to changes in
the nuclear degrees of freedom. This is equivalent to the Born-Oppenheimer
adiabatic separation between the electronic and nuclear degrees of freedom.
When M$_{\rm Q}$ is sufficiently small there will be essentially no thermal
coupling between the nuclear and electronic degrees of freedom. For
numerical convenience we choose the mass small enough to satisfy the
foregoing requirement yet large enough so that the time-step required for the
solution
of the equations of motion is not too small. The charge degrees of freedom
are to remain near zero Kelvin, since they are to be near the values which
minimize the electrostatic energy.
Within the context of molecular dynamics, this can be achieved by Nos\'{e}
thermostating\cite{Nose} the charge at a much lower temperature ($\sim$5 K)
than
the nuclei as is done in Car-Parrinello {\it ab initio} molecular
dynamics\cite{Sprik2,Bloechl,Fois}.
However, with a charge mass of 10.0 for the
TIP4P-FQ model and 11.6 (psec/e)$^2$ kcal/mole for the SPC-FQ model
and a 1 femtosecond time step for both models,
there is almost no thermal coupling
and the charge degrees-of-freedom remain at a temperature of less than 6
Kelvin
for the duration of a 50 picosecond simulation. For simulations of this
duration or shorter, no thermostating is needed to keep the charge degrees
of freedom near zero Kelvin and the atomic degrees of freedom near the
desired, much higher temperature. With this time step and M$_{\rm Q}$,
the fluc-q
models have the same energy conservation as the fixed charge models.
However, that small charge mass requires the use of a time step no greater
than about a femtosecond.
This problem could be surmounted
using multiple time scale molecular dynamics\cite{Tuckerman}.

Following Rapp\'{e} and Goddard\cite{Rappe},
the Coulomb interaction, J$_{ij}$(r), for intramolecular pairs
is taken to be the Coulomb overlap integral between Slater orbitals
centered on each atomic site,
\begin{equation}
{\rm J}_{ij}({\rm r}) =
\int d{\bf r}_i d{\bf r}_j
| \phi_{n_i}({\rm r}_i) |^2 {1 \over |{\bf r}_i - {\bf r}_j - {\bf r} |}
| \phi_{n_j}({\rm r}_j) |^2.
\label {eq:overlap}
\end{equation}
 The Slater orbitals are given by
\begin{equation}
\phi_{n_i}({\rm r}) = {\rm A}_i  {\rm r}^{n_i-1} e^{-\zeta_i r}
\label {eq:slater}
\end{equation}
and are characterized by a
principal quantum number, $n_i$, and an exponent $\zeta_i$. A$_i$ is
a normalization factor.
The value of J$_{ii}$(r) for r=0 is J$_{ii}^0$ and therefore the
value of $\zeta_i$ uniquely determines J$_{ii}^0$.
For hydrogen, $n_H$=1 and J$_{HH}^0$=${5 \over 8} \zeta_H$
and for oxygen, $n_O$=2
and J$_{OO}^0$=${93 \over 256} \zeta_O$.
The intermolecular Coulomb interaction is set equal to the
pure Coulomb interaction, 1/r, for consistency with other force fields.
The interaction J$_{\rm HH}$ is shown in Figure \ref{fig:Jr}.

Two different water geometries were used, corresponding to the commonly
used SPC\cite{BerendsenSPC} and TIP4P\cite{JorgensenKlein} water models. Both
of these models have 3 charged sites,
two positive charged hydrogen  sites and a negative charged site, and
a Lennard-Jones interaction between oxygen sites,
\begin{equation}
{\rm U}_{LJ} ({\rm r}) = 4 \epsilon \left[
\left( { \sigma \over {\rm r} } \right)^{12}
- \left( { \sigma \over {\rm r} } \right)^{6} \right]
\label {eq:lj}
\end{equation}
with a well depth,  $\epsilon$, and a diameter, $\sigma$.
The SPC potential places the negatively charged site on the oxygen
position; the TIP4P potential places this site (called the ``M-site'')
a distance of 0.15 {\AA} from the oxygen position along the dipole
direction toward the center-of-mass.
The O-H bond length and H-O-H bond angle for the potentials are
listed on Table \ref{tab:parameters}.
The TIP4P model has the
added complexity (and computational cost) of an additional interaction
site, but has the correct water geometry.
The potential energy contains the Lennard-Jones part (Eq.~(\ref{eq:lj}))
and the electrostatic part (Eq.~(\ref{eq:ec}))
and since we are defining the energies  relative to the
isolated gas-phase energy, the gas phase energy, E$_{gp}$ needs to be
subtracted.
For the isolated gas-phase water molecule, the charge constraint
gives Q$_O$ = $-$2Q$_H$ and it is straightforward to find that the
charge which minimizes the energy is
\begin{equation}
{\rm Q}_H^{gp} = { ( \tchi_O^0 - \tchi_H^0 ) \over
2 {\rm J}_{OO}^0 + {\rm J}_{HH}^0
- 4 {\rm J}_{OH}({\rm r}_{MH}) + {\rm J}_{HH} ({\rm r}_{HH})}
\label{eq:qgp}
\end{equation}
and the gas phase energy is thus
\begin{equation}
{\rm E}_{gp} = { -( \tchi_O^0 - \tchi_H^0 )^2  \over
 2 {\rm J}_{OO}^0 + {\rm J}_{HH}^0
- 4 {\rm J}_{OH}({\rm r}_{MH}) + {\rm J}_{HH} ({\rm r}_{HH})}
\label{eq:egp}
\end{equation}
where r$_{\rm MH}$ is the distance between the hydrogen and the
M-site for the TIP4P model and for the SPC model, it is the distance
between the hydrogen and the oxygen site.
We have dropped the charge independent term (E$_i$(0)),
taking this as our definition of the
zero of energy.
The total energy for N$_{molec}$ molecules is
a sum of the Lennard-Jones part, the intermolecular Coulomb part,
an intramolecular self-energy and the gas-phase energy
correction, to give
\[
{\rm E} =  \sum_{i=1}^{{\rm N}_{molec}} \sum_{j<i} \left(
4 \epsilon \left[
\left( { \sigma \over {\rm r}_{iO,jO} } \right)^{12}
- \left( { \sigma \over {\rm r}_{iO,jO} } \right)^{6} \right]
+
\sum_{\alpha=1}^3 \sum_{\beta=1}^3 {\rm Q}_{i\alpha} {\rm Q}_{j\beta}
/  {\rm r}_{i\alpha,j\beta} \right)
\]
\begin{equation}
+ \sum_{i=1}^{{\rm N}_{molec}}
\sum_{\alpha=1}^3 \left( \tchi_{\alpha}^0 {\rm Q}_{i\alpha}
+ {1\over 2} \sum_{\beta=1}^3 {\rm Q}_{i\alpha} {\rm Q}_{i\beta}
{\rm J}_{\alpha\beta}({\rm r}_{i\alpha,i\beta}) \right)
- {\rm N}_{molec} {\rm E}_{gp}
\label{eq:eij}
\end{equation}
where
${\rm r}_{i\alpha,j\beta}$ is $|$ r$_{i\alpha}$ - r$_{j\beta}$ $|$
and $\alpha$=O indicates the oxygen atom\cite{e0note}.
For periodic systems using the Ewald sum, the energy is
\[
{\rm E} =  \sum_{i=1}^{{\rm N}_{molec}} \sum_{j<i} \left(
4 \epsilon \left[
\left( { \sigma \over {\rm r}_{iO,jO} } \right)^{12}
- \left( { \sigma \over {\rm r}_{iO,jO} } \right)^{6} \right]
+
\sum_{\alpha=1}^3 \sum_{\beta=1}^3 {\rm Q}_{i\alpha} {\rm Q}_{j\beta}
{\rm erfc}(\kappa {\rm r}_{i\alpha,j\beta})/{\rm r}_{i\alpha,j\beta} \right)
\]
\[
+ {1 \over 2}
{4 \pi \over {\rm L}^3}
\sum_{{\bf G}\ne0} {1 \over {\rm G}^2} e^{-{\rm G}^2/4 \kappa^2}
\left| \sum_{i} \sum_{\alpha}
{\rm Q}_{i\alpha}
{\rm e}^{i {\bf G} \cdot {\bf r}_{i\alpha,j\beta}} \right|^2
\]
\begin{equation}
+ \sum_{i=1}^{{\rm N}_{molec}} \sum_{\alpha=1}^3 \left( \tchi_{\alpha}^0 {\rm
Q}_{i\alpha}
+ {1\over 2} \sum_{\beta=1}^3 {\rm Q}_{i\alpha} {\rm Q}_{i\beta}
({\rm J}_{\alpha\beta}({\rm r}_{i\alpha,i\beta})
- {\rm erf}(\kappa {\rm r}_{i\alpha,i\beta})/{\rm r}_{i\alpha,i\beta})
\right)
- {{\rm N}_{molec}} {\rm E}_{gp}
\label{eq:eijewald}
\end{equation}
where $\kappa$ is a screening parameter, {\bf G} is a recripocal lattice
vector of the periodic simulation cells, erf(x) is the error
function, erfc(x) is the complementary error function,
and L is the side length of the
primary simulation box\cite{Allen}.

There are three independent electrostatic parameters, $\tchi_O^0-\tchi_H^0$,
$\zeta_O$, and $\zeta_H$, since the energy is dependent on the
difference of the atomic electronegativities and the Slater exponents
describe J(r).  We have adjusted these three parameters plus the two
Lennard-Jones parameters to obtain the correct gas-phase dipole moments
and to optimize the energy, pressure,  and pair correlation functions of the
liquid.
The parameters are
given in Table \ref{tab:parameters}. In order to implement the fluc-q
procedure for the rigid bond length and rigid angle potentials used
here, the Coulomb overlap integral needs to be evaluated only at r=0
and at the intramolecular bond lengths, r$_{OH}$ and r$_{HH}$.  These
values are also given on Table \ref{tab:parameters}. This optimization
procedure does not uniquely define a set of parameters and those
listed on Table \ref{tab:parameters} are one possible good choice
which leads to improved water properties relative to the fixed charge
models, as discussed in the next section.
The electrostatic parameters, $\tchi_{\rm O}^O-\tchi_{\rm H}^0$,
J$_{\rm HH}^0$ and
J$_{\rm OO}^0$ are within the range of values used in previous EE
models\cite{Allinger,Mortier,No,Rappe,Richards}.

\section{Numerical Method}
\setcounter{equation}{0}
\label{sec:numerical}

The molecular dynamics
simulations were performed on the Connection Machine CM-5
with 256 molecules. Periodic
boundary conditions were imposed, using the Ewald sum for the long-ranged
electrostatic potentials. The screening parameter, $\kappa$, was set
to 5/L and 256
reciprocal lattice vectors were used in the Fourier
space sum. A time step of 1 femtosecond and the SHAKE algorithm for
enforcing bond constraints were used{\cite{Allen}}.
The data reported in the next section is from 20 separate
50 picosecond runs for each FQ model.
MD is implemented on the parallel architecture of the CM-5 by
arranging data on a 2-dimensional grid, so that
each virtual node ($i,j$)
contains all the information (namely
r$_{i\alpha,j\beta}$ and Q$_{i\alpha}\cdot$Q$_{j\beta}$)
for the interactions between
molecules $i$ and $j$\cite{Lynch}.
This is done as follows:
\begin{enumerate}
 \item For $\alpha$=1 to $\alpha$=N$_{atoms}$,
{\bf r}$_{i\alpha}$ is placed on the first row of
an N$_{molec}\times$N$_{molec}$ array
({\bf A}$_{\alpha}$(i,1)={\bf r}$_{i\alpha}$)
and the data is then spread through each row of
the array using the parallel copy operation
(this step takes 3\% of one time step).
\item Similarly,
{\bf r}$_{i\alpha}$ is placed on the first column of a
different matrix ({\bf B}$_{\alpha}$(1,i)={\bf r}$_{i\alpha}$)
and spread though each column(3\%).
\item Nearest image r$_{i\alpha,j\beta}$ values
can now be computed from {\bf A}$_{\alpha}$(i,j) and {\bf B}$_{\beta}$(i,j)
without any communication between different virtual nodes (14\%).
\item A similar ``spread-spread'' algorithm is used to place
Q$_{i\alpha}$ and Q$_{j\beta}$
on each virtual node $i,j$ (3\%) and
Q$_{i\alpha}\cdot$ Q$_{j\beta}$ is calculated (1\%).
\item All pairwise forces and potential energy terms can be calculated on
each virtual node without communication (the Lennard-Jones
term takes about 3\%, the real space Ewald term takes 33\%, and the Coulomb
self-term takes 3\% of a time step).
The Fourier part of the Ewald sum (see Eq.~\ref{eq:eijewald})
does not involve pair terms and is calculated separately from the
N$_{molec}^2$ process (21\%).
\item The total energy and the force on atom $i$ from all other atoms $j$
are found from a parallel add operation across the virtual nodes (10\%).
\item The positions and charges are propagated using Eqs.~\ref{eq:newton} and
\ref{eq:cp}(0\%), bond constraints are enforced using SHAKE (4\%),
 and the steps are repeated.
\end{enumerate}
On a 16 processor CM-5, we found performances of about
0.4 seconds/time step, which is about a factor of 10 faster than a
comparable program on a IBM 580\cite{cmnote}.
The CPU required for the fluctuating charge model is only a factor
of 1.10 larger than for the corresponding fixed-charge model.

\section{Results}
\setcounter{equation}{0}
\label{sec:results}
The properties of the fixed-charge and fluc-q models for the water
monomer, water dimer, and liquid water are listed in
Tables \ref{tab:propertiest} and \ref{tab:propertiess}.
Our reported error bars represent two standard deviation error
estimates.
Table \ref{tab:propertiest} lists the results for the TIP4P-FQ
model, comparing to other TIP4P geometry models: TIP4P\cite{JorgensenKlein},
the Watanabe-Klein (WK) model\cite{Watanabe}, and the dipole polarizable
SRWK (SRWK-P) model\cite{Sprik}.
The SRWK model has a slightly different geometry than TIP4P, the
length r$_{\rm OM}$ is 0.26{\AA} rather than 0.15{\AA}.
Table \ref{tab:propertiess} lists the results for some of the many
models with an SPC geometry, including SPC\cite{BerendsenSPC}, SPC/E
\cite{Berendsen}, polarizable SPC (PSPC)\cite{Ahlstrom}, and
flexible charge SPC (SPC-FQ).
The TIP4P, WK, SPC, and SPC/E models are all
non-polarizable fixed-charge models, the WK and SPC/E models include
a correction for the polarization energy, discused below.
The WK model also includes a correction for the quantum librational energy.

For the monomer, the electrostatic parameters are chosen to give the correct
gas phase dipole moment.
Another property of the isolated molecule is
the dipole polarizability tensor, \boldmath $\alpha$\unboldmath, defined by
\begin{equation}
\mbox{\boldmath{$\mu$}}^{ind} =
\mbox{\boldmath{$\alpha$}} \cdot {\bf  E}
\label {eq:dipole}
\end{equation}
where \boldmath${\mu}^{ind}$\unboldmath is the dipole moment induced by the
external electric field, {\bf E}.
\boldmath${\mu}^{ind}$\unboldmath can be determined by adding a term
$-\mbox{\boldmath{$\mu$}} \cdot {\bf  E}$ to Eq.~\ref{eq:ec}
for the monomer and minimizing the total potential energy with
respect to the charge.
If the plane of the molecule is in the zy plane and the dipole (C$_2$-axis)
is along the z-direction, then it is found that
\[
\alpha_{zz} = { 2 {\rm z}_{MH}^2 \over
 2 {\rm J}_{OO}^0 + {\rm J}_{HH}^0
- 4 {\rm J}_{OH}({\rm r}_{MH}) + {\rm J}_{HH} ({\rm r}_{HH})}
\]
\[
\alpha_{yy} = { {\rm r}_{HH}^2 /2
\over {\rm J}_{HH}^0 - {\rm J}_{HH} ({\rm r}_{HH}) }
\]
\begin{equation}
\alpha_{xx} = 0
\label{eq:alphas}
\end{equation}
where z$_{MH}$ is the z-component of the
distance from the negative charge site (the M site
for TIP4P, the oxygen site for SPC) to the positive charge sites and
$\alpha_{xx}$ is zero since all the charges are in the zy plane and
no dipole induction is possible out of plane.
Experimentally, $\alpha$ is almost isotropic\cite{Murphy}, so the
lack of polarizability in the x-direction is clearly a deficiency
in fluc-q models, but corrections are possible\cite{Dinar}.

The dimer properties listed in Table \ref{tab:propertiest}
and \ref{tab:propertiess} are the energy of the minimum energy
configuration and the oxygen-oxygen distance of this configuration. Pair
potentials, such as SPC and TIP4P, are parametrized to give the measured
liquid state energies and radial distribution functions. It is known that
this parametrization of these pairwise interaction models overestimates the
gas phase water dimer energy.
The fluctuating charge potentials
predict an oxygen-oxygen separation closer to the experimental value
but underestimate the dimer energy.

We have calculated both static and dynamical properties of liquid state
water at a temperature $T=298$K and $\rho= 1$g/cm$^3$.
The error bars represent two standard deviations.
The parameters for
both fluc-q models are chosen to give a binding energy of -9.9 kcal/mole.
This energy, unlike the fixed-charge potential energies, includes the self
polarization contribution arising from the difference in the
internal energy given by Eq.~\ref{eq:expan} for the liquid state
charges and the gas phase charges,
\[
\Delta {\rm E}_{{\rm self~pol}} =
\sum_{\alpha=1}^3 \left[ \langle {\rm E}_{\alpha}({\rm Q}_{\alpha}^{liq})
\rangle - {\rm E}_{\alpha}({\rm Q}_{\alpha}^{gas}) \right].
\]
This self polarization energy is the difference
between the self-energy in the liquid phase and the
gas phase Coulombic energy.
The average self
polarization energy is 5.7 kcal/mole for TIP4P-FQ and 7.6 kcal/mole for
SPC-FQ, which represents a large contribution to the total energy. The
dipole polarizable model of Sprik and Klein has a polarization energy of
5.9 kcal/mole. It has been noted that because the polarization energy is
ignored in the parametrization of the SPC and TIP4P interaction potentials,
these models underestimate the attractive pair interactions in
water\cite{Berendsen}. One simple correction for this is to subtract an
estimate of the polarization energy, $(\mu_{liq}-\mu_{gas})^2/2\alpha$,
from the experimentally measured binding energy and to use the result to
parameterize the pair potential. Here $\mu_{liq}$ and $\mu_{gas}$ are
respectively the liquid state and gas phase permanent dipole moments used
in the models. From this, it follows that the strength of the pairwise
interaction must be increased to give the correct energy of $-10$ kcal/mole.
This has been done for the SPC model (giving the SPE/E model
\cite{Berendsen}) and TIP4P (giving the the WK model \cite{Watanabe}), both
of which have increased atomic charges and reduced diffusion constants
compared to SPC or TIP4P.

The pair correlation functions give detailed information about the
structure of the liquid. The TIP4P and TIP4P-FQ oxygen-oxygen radial
distribution functions, g$_{\rm OO}$(r), are shown in
Fig.~\ref{fig:gootip4p}, and are compared to the neutron diffraction results
of Soper and Phillips\cite{SoperPhillips}.
Recent experiments of Soper and Turner
indicate that there is a large experimental uncertainty in the
peak heights of the pair correlation functions, perhaps due to the
use of different methods for removing the
contribution from self or single atom scattering\cite{SoperTurner}.
The peak positions show much less uncertainty and therefore
provide more reliable points for comparison.
The g$_{\rm OO}$(r) of the flexible charge model has a first peak
at a larger r than the fixed charge model and shows more long-ranged
ordering due to the increased charges.
The oxygen-hydrogen
(Fig.~\ref{fig:gohtip4p}) and hydrogen-hydrogen radial distribution function
(Fig.~\ref{fig:ghhtip4p}) for TIP4P-FQ and TIP4P models are also shown. The
radial distribution functions for the SPC-FQ potential are shown in
Figs.~\ref{fig:goospc},\ref{fig:gohspc}, and \ref{fig:ghhspc}.
Again, the flexible charge g$_{\rm OO}$(r) shows more long-range correlation
than the corresponding fixed charge model, but, in general,
the SPC-FQ model does not
give as accurate pair correlation functions as the TIP4P-FQ model.
The number of
nearest neighbors can be determined by integrating g$_{\rm OO}$(r) over the
first peak. Experimentally, the first minimum occurs at 3.3 {\AA} and using
this value as the limit of the first peak gives essentially
the same coordination
number for all models: 4.2 (SPC-FQ), 4.3 (TIP4P), and 4.4 (TIP4P-FQ, experiment
and SPC).

The average dipole moment, $\langle \mu \rangle$, is increased for both
fluctuating charge
models over the corresponding fixed-charge models. The value
of the liquid state dipole moment is not known experimentally.
The experimental value for ice is 2.6 D\cite{CoulsonEisenberg}.
Theoretical studies which use potentials which have the correct dipole
polarizability and quadrupole moments find a dipole moment of 2.5 D
\cite{Barnes} and 2.45 to 2.7 D, depending on the details of the
non-electrostatic part of the potential \cite{Patey}. Additionally, it has
been observed that the dependence of the dielectric constant on the dipole
moment is such that to have a dielectric constant close to 80, the
potential must have a dipole moment in the range of 2.3 to 2.6
D\cite{Sprik}.
The distributions of the dipole moment are
shown in Figure \ref{fig:pmu}. The full width at half maximum is 0.42
for the SPC-FQ and 0.49 for TIP4P-FQ. The SPC geometry, with greater
distances between the charge sites, is more polarizable and therefore the
distribution of the $|\mu|$ is broader. In the fluctuating charge models
the instantaneous dipole moment does not always lie along the C$_2$
direction (what we have previously defined as the z-axis).
The average of the component of the dipole moment along the C$_2$
axis, $\mu_{\rm z}$,
for the TIP4P-FQ model is 2.59 D, which when compared to the total dipole
moment of 2.62 D indicates that there are small fluctuations of the
dipole moment away from the C$_2$ axis.
For the SPC-FQ model, $\mu_{\rm z}$ is 2.81 D and the total dipole moment
is 2.83 D.

The static
dielectric constant, $\epsilon_0$, for the FQ potentials, calculated from
the fluctuations in the total dipole of the central
simulation box, {\bf M}, according to\cite{Allen}
\begin{equation}
\epsilon_0 = \epsilon_{\infty} + \left( {4 \pi \rho \over 3 k T} \right) \left[
{ \langle {\bf M}^2 \rangle - \langle {\bf M} \rangle^2 \over
{\rm N}_{molec} } \right]
\label{eq:diel}
\end{equation}
is 79$\pm$8 for TIP4P-FQ and 116$\pm$18 for SPC-FQ.
Eq. \ref{eq:diel} was evaluated from 1 nanosecond run.
The dielectric constant of the TIP4P-FQ model is in good agreement with
experiment, which is consistent with earlier findings that models with
a dipole moment of 2.6 D have a dielectric constant near
80\cite{Patey,Watanabe,Sprik}. The SPC-FQ model is in poorer agreement due to
the fact that
the liquid state dipole moment is large.
A large dipole moment\cite{Ahlstrom,BernardoLevy,Dang} and dielectric constant
\cite{SmithDang} is also seen in dipole
polarizable SPC models.
{}From this perspective the TIP4P geometry is better than SPC.

The fluctuating charge models have a dielectric response
characterized by an infinite frequency dielectric constant,
$\epsilon_{\infty}$. Neumann and Steinhauser have derived expressions
for calculating $\epsilon_{\infty}$ for dipole polarizable
models\cite{NeumannSteinhauser}.
Expressions for $\epsilon_{\infty}$ for fluctuating charge models are
similar.
The Coulomb energy can be written as
\begin{equation}
{\rm U}({\bf Q}) = (\tchi_{\rm H}^0 - \tchi_{\rm O}^0) {\bf Q} \cdot
{\bf I} + {1 \over 2} {\bf Q} \cdot {\bf J} \cdot {\bf Q}
\end{equation}
where {\bf Q} is a 2 N$_{\rm molec}$ dimensional vector containing
the hydrogen charges (the oxygen charges are eliminated from the
equation using the charge neutrality constraint), {\bf I} is the identity
vector, and {\bf J} is a (2N$_{\rm molec} \times 2{\rm N}_{\rm molec}$) matrix.
The elements of {\bf J} are given by
\begin{equation}
{\rm J}_{i\alpha,j\beta}=
\left\{ \begin{array}{ll}
1/{\rm r}_{i\alpha,j\beta} - 1/{\rm r}_{i\alpha,jM}
-1/{\rm r}_{iM,j\beta} + 1/{\rm r}_{iM,jM}  & i\neq j \\
{\rm J}_{\alpha\beta}({\rm r}_{i\alpha,j\beta})
-{\rm J}_{\alpha O}({\rm r}_{i\alpha,jM})
-{\rm J}_{O\beta}({\rm r}_{iM,j\beta})
+{\rm J}_{OO}({\rm r}_{iM,jM}) & i=j
\end{array}. \right.
\label{eq:jmatrix}
\end{equation}
With the Ewald sum---which we are using in the present calculations---for the
long range Coulomb interaction, each 1/r and J$_{\alpha\beta}$
interaction in Eq.~\ref{eq:jmatrix}
is replaced as follows,\cite{Allen}
\[
1/{\rm r}_{i\alpha,j\beta} \Rightarrow {\rm erfc}(\kappa {\rm
r}_{i\alpha,j\beta})/{\rm r}_{i\alpha,j\beta}
+ {4 \pi \over {\rm L}^3 } \sum_{{\bf G}\ne 0}{1 \over {\rm G}^2} e^{-{\rm
G}^2/4\kappa^2}
{\rm cos}({\bf G} \cdot {\bf r}_{i\alpha,j\beta})
\]
\[
{\rm J}_{\alpha,\beta}({\rm r}_{i\alpha,i\beta}) \Rightarrow
{\rm J}_{\alpha,\beta}({\rm r}_{i\alpha,i\beta})
-{\rm erf}(\kappa {\rm r}_{i\alpha,i\beta})/{\rm r}_{i\alpha,i\beta} +
{4 \pi \over {\rm L}^3 } \sum_{{\bf G}\ne 0}{1 \over {\rm G}^2} e^{-{\rm
G}^2/4\kappa^2}
{\rm cos}({\bf G} \cdot {\bf r}_{i\alpha,i\beta}).
\]
The minimum energy charges are given by
\begin{equation}
{\bf Q}  = - (\tchi_{\rm H}^0 - \tchi_{\rm O}^0) {\bf J}^{-1} \cdot {\bf I}
\end{equation}
where ${\bf J}^{-1}$ is the inverse of {\bf J}. In the presence
of a spatially homogeneous external electric field, {\bf E}, the
charges are
\begin{equation}
{\bf Q} = {\bf Q}^0 + {\bf J}^{-1} \cdot \delta {\bf r} \cdot {\bf E}
\end{equation}
where ${\bf Q}^0$ are the charges in the absence of the field and
$\delta {\bf r}_{i\alpha}$=${\bf r}_{i\alpha}-{\bf r}_{iO}$.
The energy is
\begin{equation}
{\rm U}= {\rm U}^0 - {\bf M} \cdot {\bf E} - {1 \over 2} {\bf E}
\cdot {\bf A} \cdot {\bf E}
\label{eq:ufield}
\end{equation}
where U$^0$ is the energy in the absence of the field and {\bf A}
is the polarizability of the system, given by
\begin{equation}
{\rm A}_{i\alpha,j\beta} =
\delta {\bf r}_{i\alpha} \cdot \delta {\bf r}_{j\beta}
{\rm J}_{i\alpha,j\beta}^{-1}.
\end{equation}
Following Ref. \cite{NeumannSteinhauser}, $\epsilon_{\infty}$ is
\begin{equation}
\epsilon_{\infty} = 1 + {4 \pi \over 3 {\rm V} }
\sum_{i=1}^{N_{\rm molec}} \sum_{j=1}^{N_{\rm molec}}
\sum_{\alpha=2}^3 \sum_{\beta=2}^3
\langle {\rm A}_{i\alpha,j\beta} \rangle.
\label{eq:einfty}
\end{equation}
Eq. \ref{eq:einfty} was evaluated
every 50 picoseconds in the course of the simulations, which
is a frequent enough sampling rate to provide a precise estimate of
$\langle {\rm A}_{i\alpha,j\beta} \rangle$.
The value obtained for $\epsilon_{\infty}$ from
both fluctuating charge models is about 1.6, close
to the experimental value of 1.79 \cite{Buckingham}.
$\epsilon_{\infty}$ is underestimated
because the perpendicular polarizability,
$\alpha_{\rm xx}$, is zero.
To leading order in $\alpha$, and thus $\epsilon_{\infty}$-1,
the total system polarizability, {\bf A},
is proportional to
Tr(\boldmath${\alpha}$\unboldmath)\cite{NeumannSteinhauser}.
It then follows that to order $\alpha$,
\begin{equation}
{ \epsilon_{\infty}({\rm FQ})-1  \over
\epsilon_{\infty}({\rm exact})-1 }
\approx
{ {\rm Tr}[\mbox{\boldmath{$\alpha$}}({\rm FQ})] \over
  {\rm Tr}[\mbox{\boldmath{$\alpha$}}({\rm exact})] }
\end{equation}
which is true for the data from
Tables \ref{tab:propertiest} and \ref{tab:propertiess}.
Therefore, $\alpha_{\rm xx}=0$ produces
an $\epsilon_{\infty}$ smaller than the experimental value.

The frequency dependent dielectric constant can be calculated
from\cite{NeumannSteinhauser}
\begin{equation}
{\epsilon(\omega) - \epsilon_{\infty} \over
\epsilon_0 - \epsilon_{\infty} } =
1 - i \omega {\cal L}_{i\omega} [\phi(t)]
\label{eq:epsomega}
\end{equation}
where ${\cal L}_{i\omega}$ denotes the Laplace operator and
$\phi(t)$ is the normalized
time autocorrelation function of the system's
total dipole ({\bf M}=$\sum$
\boldmath${\mu}$\unboldmath$_i$),
\begin{equation}
\phi(t) = \langle {\bf M}(t) \cdot {\bf M}(0) \rangle / \langle {\bf M}^2
\rangle .
\end{equation}
$\phi$(t) has a short-time oscillatory part, due to librational
motions of the hydrogen atoms. At long times, $\phi$(t) decays
exponentially and the decay constant is the Debye relaxation time,
$\tau_{\rm D}$ (see Fig.~\ref{fig:mut}).
In order to perform the Laplace transform to get
$\epsilon(\omega)$, we set $\phi$(t)=Aexp[-t/$\tau_{\rm D}$] for
times longer than 0.5 picoseconds. The parameters A and $\tau_{\rm D}$
were chosen to give a smooth interpolation between the calculated
$\phi$(t) and the exponential fit (see Fig.~\ref{fig:mut}).
The frequency dependent dielectric constant for the TIP4P-FQ model
is shown in Fig.~\ref{fig:epst}. The agreement with the
experimental results\cite{Kaatze,Afsar,Rusk} is very good.
The close agreement in the low-frequency microwave range is due
to the fact that the model gives
accurate values of $\epsilon_0$ and $\tau_{\rm D}$. The features
at frequencies higher than 300 psec$^{-1}$ are due to bond stretches
and bends and so are not present in the rigid geometry models
used here. The highest frequency feature given by the TIP4P-FQ model
is the librational mode, which shows a peak in $\epsilon^{\prime\prime}$
at 130 psec$^{-1}$. The experimental peak is at 90 psec$^{-1}$\cite{Rusk}.
Another notable feature is at 25 psec$^{-1}$ which has been interpreted as
a translational vibration of a water molecule in its cage of
nearest neighbors\cite{Walrafen}. This feature is not present in the
spectrum for non-polarizable water models such as
TIP4P\cite{Neumann} or MCY\cite{Neumann2}.
As argued by Neumann, this translational motion will not change the
system's dipole moment much for non-polarizable models, but for
polarizable models, the translation motion will induce a change
in the dipole moment\cite{Neumann}.
Therefore this feature can only be seen in polarizable models.
The fluctuating charge models
support that argument.
The frequency dependent dielectric constant for SPC-FQ is shown on
Fig.~\ref{fig:epss}.
The agreement with experiment is not as good as the TIP4P-FQ model,
primarily because the static dielectric constant is overestimated.
The librational peak is at 160 psec$^{-1}$ for the SPC-FQ
model.

Lastly, we  examine the dynamical properties of the fixed and fluc-q
model potentials.
In general, the flexible charge models have slower translational
and rotational time-scales than the fixed-charge models, primarily
due to the stronger electrostatic interactions from the higher
charges.
The translational diffusion constant, D, is determined from the
Einstein relation
\begin{equation}
{\rm D} = \lim_{t \rightarrow \infty} {1 \over 6 t }
\langle | {\bf r}_i^{\rm CM} (t) - {\bf r}_i^{\rm CM} (0) |^2 \rangle
\label{eq:einstein}
\end{equation}
where ${\bf r}_i^{\rm CM}$(t) is the position of the center-of-mass
of molecule $i$ at time t.
The diffusion constants for the flexible charge models are smaller
than the fixed-charge models and closer to the experimental
value (see Tables \ref{tab:propertiest} and \ref{tab:propertiess}).
Rotational time constants are calculated from
\begin{equation}
{\rm C}_l^{\alpha}(t) = \langle {\rm P}_l ( {\bf e}_i^{\alpha}(t) \cdot
{\bf e}_i^{\alpha}(0) ) \rangle
\label {eq:rotation}
\end{equation}
where P$_l$ is a Legendre polynomial and ${\bf e}_i^{\alpha}$
are unit vectors along molecule $i$'s principle axis of rotation\cite{Impey}.
Rotations around the axis connecting the hydrogen atoms (the y-axis)
can be measured by proton NMR.
The zero frequency component of the Fourier transform of C$_2^y$(t)
gives the NMR relaxation time, $\tau_{NMR}$.
The correlation function C$_2^y$(t)
has a long range exponential decay, given by A$_2^y$ exp(-t/$\tau_2^y$),
and a short range, oscillatory part (see Fig.~\ref{fig:c2y}).
The zero frequency part of the Fourier transform
of C$_2^y$(t) is, to a good approximation, given by  A$_2^y\tau_2^y$, since
the short range part will not contribute much to the transform.
The flexible charge models exhibit slower reorientational dynamics
than the fixed-charge models and
are in closer agreement with the experimental value\cite{Jonas}.

\section{Conclusions}
\setcounter{equation}{0}
\label{sec:conclusions}

The fluctuating charge (fluc-q) water models, using either
a TIP4P or SPC geometry, were shown to give important improvements
over fixed charge models.
The electronic properties of the
fluc-q models are such that in the gas phase they give the correct
dipole moment (this is by construction) and in the liquid phase,
the dielectric properties are well reproduced for a range of
frequencies (see Figs.~\ref{fig:epst} and \ref{fig:epss}).
The dielectric properties are best for the TIP4P-FQ model which
gives a static dielectric constant, $\epsilon_0$,
an infinite frequency dielectric constant, $\epsilon_{\infty}$, and
a Debye relaxation time, $\tau_D$, close to the experimental values.
The SPC-FQ model overestimates $\epsilon_0$,
but $\epsilon_{\infty}$ and $\tau_D$ are accurate.
The fluc-q models also show (at 25 psec$^{-1}$)
a feature in the dielectric spectrum
originating from translational motion
of a water molecule in the cage of its neighbors. This is a feature
which fixed charge models do not show\cite{Neumann,Neumann2}.
Translations will have a large effect on the system's dipole moment
only for polarizable models, so this feature of $\epsilon$($\omega$)
is an indication of the coupling between the electronic and
nuclear degrees-of-freedom.

In addition, the fluc-q models give good estimates for the
liquid-state radial distribution functions (see Figs.~\ref{fig:gootip4p}-
\ref{fig:ghhspc}) and dynamical properties such as the
diffusion constant and $\tau_{\rm NMR}$ (see Tables \ref{tab:propertiest}
and \ref{tab:propertiess}).
Since the charges are not fixed to values which represent mean field
values for a particular single-component phase,
this method should be transferable to
studies of heterogeneous
systems in which deviations from the mean
field charge values should be greater than in pure systems\cite{Barnes}.
The fluc-q method assigns two parameters to each element
corresponding to the two terms in the energy expansion
(Eq. \ref{eq:expan}). The first order term is
the Mullikan electronegativity, $\tchi$, and
the second order term is determined by a Slater exponent,
$\zeta$, through the Coulomb overlap integral (Eq. \ref{eq:overlap}).
Extensions to more complex molecules would require additional
terms for other elements, perhaps to be taken from other
electronegativity equalization schemes
\cite{Allinger,Mortier,No,Rappe,Richards}.

All in all, the fluc-q water models are as successful as the best of the
dipole polarizable models,
such as SRWK-P\cite{Sprik}, RER\cite{WallqvistBerne}, and RPOL\cite{Dang}.
The dipole polarizable models introduce an interaction
(the dipole-dipole interaction)
not present in fixed-charge models, which, together with solving
for the induced dipole moments, increases the CPU time by about
a factor of 2 over fixed charge models\cite{WallqvistBerne,VanBelle3}.
The fluc-q method
introduces no new interactions
and the propagation of the charges using extended
Lagrangian methods increases the computational cost by only a modest amount
(about 1.1). The dynamical fluctuating charge method
is therefore a tractable means for the inclusion of charge polarization
effects and will be useful for the study of large systems,
such as proteins in aqueous environments.

\vspace {30pt}

\noindent \large {\bf{Acknowledgements}}
\normalsize
\vspace {10pt}

This work was funded by a grant from the National Institutes of Health
(GM 43340)
and was done on the Thinking Machines CM-5 in the NIH
Biotechnology Resource Center at Columbia University.
SJS was supported by a National Defense Science and Engineering
Graduate Fellowship (DAAL 03-91-G-0277).

\clearpage
\baselineskip=24pt

\bibliography{h2oeecp}
\clearpage

\begin{table}
\caption{\baselineskip=24pt Potential parameters for the fixed-charge
potentials, SPC and
TIP4P,
and the flexible charge SPC and TIP4P (SPC-FQ, TIP4P-FQ) models.
The last four terms, J$_{\rm AB}$, are determined by $\zeta_{\rm H}$
and $\zeta_{\rm O}$ and so are not independent parameters.}
\vspace{5 mm}
\begin{tabular}{lllll}
 & SPC$^a$ & TIP4P$^b$ & SPC-FQ & TIP4P-FQ \\
\cline{1-5}
$\epsilon$ (kcal/mole) & 0.1554 & 0.1550 & 0.2941 & 0.2862 \\
$\sigma$ ({\AA}) & 3.166 & 3.154 & 3.176 & 3.159 \\
$\theta_{HOH}$ (degrees) & 109.47 & 104.52 & 109.47 & 104.52 \\
r$_{OH}$ ({\AA}) & 1.0 & 0.9572 & 1.0 & 0.9572 \\
r$_{OM}$ ({\AA}) & 0.0 & 0.15 & 0.0 & 0.15 \\
Q$_H$ (e) & 0.41 & 0.52 & & \\
$\tchi_O - \tchi_H$ (kcal/mole e) & & & 73.33 & 68.49 \\
$\zeta_O$ (a$_0^{-1}$) & & & 1.61 & 1.63 \\
$\zeta_H$ (a$_0^{-1}$) & & & 1.00 & 0.90 \\
J$_{OO}^0$ (kcal/(mole e$^2$)) & & &367.0 & 371.6 \\
J$_{HH}^0$ (kcal/(mole e$^2$)) & & &392.2& 353.0 \\
J$_{OH}$(r$_{OH}$) (kcal/(mole e$^2$)) & & &276.0 & 286.4 \\
J$_{HH}$(r$_{HH}$) (kcal/(mole e$^2$)) & & &196.0 & 203.6 \\
\cline{1-5}
\end{tabular}
\label{tab:parameters}
\end{table}
\noindent
a) Ref. \cite{BerendsenSPC},
b) Ref. \cite{JorgensenKlein}.

\clearpage
\newpage

\begin{table}
\caption{\baselineskip=24pt Properties for potentials with the TIP4P
geometry: the fixed-charge TIP4P
and Watanabe-Klein (WK) models,
the SRWK dipole polarizable model (SRWK-P), and the
flexible charge model (TIP4P-FQ). Properties listed are the
the gas-phase dipole moment, the dipole polarizabilities, $\alpha_{ii}$
(the y and z directions lie in the plane of the molecule, with the
z-axis along the C$_2$ axis), the energy of the dimer in its minimum
energy configuration, the distance between oxygen atoms for the
minimum dimer configuration, and properties of the liquid as indicated.}

\vspace{6 mm}
\begin{tabular}{llllll}
 & TIP4P$^a$ & WK$^b$ & SRWK-P$^c$ & TIP4P-FQ & experimental \\
\cline{1-6}
Gas-phase dipole moment (Debye) & 2.18 & 2.60 & 1.85 & 1.85& 1.85$^d$ \\
$\alpha_{zz}$ ({\AA}$^3$) & 0 & 0 & 1.44 & 0.82 & 1.468$\pm$0.003$^e$\\
$\alpha_{yy}$ ({\AA}$^3$) & 0 & 0 & 1.44 & 2.55 & 1.528$\pm$0.013$^e$\\
$\alpha_{xx}$ ({\AA}$^3$) & 0 & 0 & 1.44 & 0    & 1.415$\pm$0.013$^e$\\
Dimer energy (kcal/mole) & -6.3 & & & -4.5 & -5.4$\pm$0.7$^f$\\
Dimer O-O length ({\AA}) & 2.75 &  & & 2.92 & 2.98$^f$ \\
\multicolumn{3}{l} {Liquid state properties (T=298 K, $\rho$=1.0 g/cm$^3$)}   &
     &     &  \\
$\; \; \;$ Energy (kcal/mole)   & -10.1$^a$ &-10.2$^b$  & -11.1$^c$ &-9.9 &
-9.9$^a$ \\
$\; \; \;$ Pressure (kbar)      & 0.0$^b$   & 0.1$^b$  &  0.6$^c$ &
-0.16$\pm$0.03 & 0.0 \\
$\; \; \;$ Dipole moment (Debye)& 2.18 & 2.60$^b$ & 2.63$^c$ & 2.62&     \\
$\; \; \;$ $\epsilon_0$ & 53$\pm$2$^g$ &80$\pm$8$^b$ &86$\pm$10$^c$&79$\pm$8  &
78$^h$ \\
$\; \; \;$ $\epsilon_{\infty}$ & 1 &1  &  & 1.592$\pm$0.003 & 1.79$^h$\\
$\; \; \;$ Diffusion constant(10$^{-9}$ m$^2$/s) &  3.6$\pm$0.2$^b$  &
1.1$\pm$0.3$^b$& 2.4$\pm$0.3$^c$  & 1.9$\pm$0.1 & 2.30$^i$  \\
$\; \; \;$ $\tau_{NMR}$ (ps)    & 1.4$\pm$0.2$^b$  & 3.8$\pm$0.3$^b$ &
   & 2.1$\pm$0.1 & 2.1$^j$  \\
$\; \; \;$ $\tau_{D}$ (ps)    & 7$\pm$2$^b$ & 22$\pm$4$^b$ & &  8$\pm$2 &
8.27$\pm$0.02$^k$  \\
\cline{1-6}
\end{tabular}
\label{tab:propertiest}
\vspace{5mm}

\noindent
a) Ref. \cite{JorgensenKlein},
b) Ref. \cite{Watanabe},
c) Ref. \cite{Sprik},
d) Ref. \cite{Shepard},
e) Ref. \cite{Murphy},
f) Ref. \cite{Odutola},
g) Ref. \cite{Neumann},
h) Ref. \cite{Buckingham},
i) Ref. \cite{Krynicki},
j) Ref. \cite{Jonas},
k) Ref. \cite{Kaatze}
\end{table}

\clearpage

\begin{table}
\caption{\baselineskip=24pt Properties for potentials with the SPC
geometry: the fixed-charge SPC and SPC/E models,
dipole polarizable model (PSPC), and the
flexible charge model (SPC-FQ). Properties listed are the
the gas-phase dipole moment, the dipole polarizabilities, $\alpha_{ii}$
(the y and z directions lie in the plane of the molecule, with the
z-axis along the C$_2$ axis), the energy of the dimer in its minimum
energy configuration, the distance between oxygen atoms for the
minimum dimer configuration, and properties of the liquid as indicated.}

\vspace{6 mm}
\begin{tabular}{llllll}
 & SPC$^a$ & SPC/E$^b$ & PSPC$^c$ & SPC-FQ & experimental \\
\cline{1-6}
Gas-phase dipole moment (Debye) & 2.27 & 2.35 & 1.85 & 1.85& 1.85$^d$ \\
$\alpha_{zz}$ ({\AA}$^3$) & 0 & 0 & 1.44 & 1.02 & 1.468$\pm$0.003$^e$\\
$\alpha_{yy}$ ({\AA}$^3$) & 0 & 0 & 1.44 & 2.26 & 1.528$\pm$0.013$^e$\\
$\alpha_{xx}$ ({\AA}$^3$) & 0 & 0 & 1.44 & 0    & 1.415$\pm$0.013$^e$\\
Dimer energy (kcal/mole) & -6.7 & & &  -4.4  & -5.4$\pm$0.7$^f$\\
Dimer O-O length ({\AA}) & 2.75 &  & &  2.94 & 2.98$^f$ \\
\multicolumn{3}{l} {Liquid state properties (T=298 K, $\rho$=1.0 g/cm$^3$)}   &
     &     &  \\
$\; \; \;$ Energy (kcal/mole)   & -10.0$^g$ &-9.9$^h$  & -9.1$^c$ &-9.9 &
-9.9$^i$ \\
$\; \; \;$ Pressure (kbar)      & 0.3$^g$   & -0.08$\pm$0.04$^h$  &  &
0.03$\pm$0.05 & 0.0 \\
$\; \; \;$ Dipole moment (Debye)& 2.27 & 2.35 & 2.9$^c$  & 2.83  &     \\
$\; \; \;$ $\epsilon_0$ &68$\pm$7$^j$  & 67$\pm$10$^h$& &116$\pm$18 & 78$^k$ \\
$\; \; \;$ $\epsilon_{\infty}$ & 1 &1  & &1.606$\pm$0.002 & 1.79$^k$ \\
$\; \; \;$ Diffusion constant(10$^{-9}$ m$^2$/s) &  3.3$\pm$0.2$^g$  &
2.4$\pm$0.4$^h$& 2.0$\pm$0.2$^h$  & 1.7$\pm$0.1 & 2.30$^l$  \\
$\; \; \;$ $\tau_{NMR}$ (ps)    & 1.1$\pm$0.2$^g$  & 1.9$\pm$0.1$^h$ &     &
2.2$\pm$0.1    & 2.1$^m$  \\
$\; \; \;$ $\tau_{D}$ (ps)    & 11$\pm$2$^g$ & 10$\pm$3$^h$ &     & 9$\pm$3  &
8.27$\pm$0.02$^n$  \\
\cline{1-6}
\end{tabular}
\label{tab:propertiess}
\vspace{5mm}

\noindent
a) Ref. \cite{BerendsenSPC},
b) Ref. \cite{Berendsen},
c) Ref. \cite{Ahlstrom},
d) Ref. \cite{Shepard},
e) Ref. \cite{Murphy},
f) Ref. \cite{Odutola},
g) Ref. \cite{Watanabe},
h) Ref. \cite{SmithDang},
i) Ref. \cite{JorgensenKlein},
j) Ref. \cite{Alper},
k) Ref. \cite{Buckingham},
l) Ref. \cite{Krynicki},
m) Ref. \cite{Jonas},
n) Ref. \cite{Kaatze}
\end{table}

\clearpage

\begin{figure}[ht]
\include{h2oeecpfig1}
\caption{Coulomb interaction for H-H pairs, comparing the
intramolecular Coulomb overlap interaction (solid line) with the
intermolecular pure Coulomb, 1/r, interaction (dotted line),
in kcal/mole/e$^2$.}
\label{fig:Jr}
\end{figure}

\begin{figure}[ht]
\include{gootip4p}
\caption{Oxygen-oxygen radial distribution function for the
TIP4P-FQ (solid line) and TIP4P (dotted line) potentials, comparing
to the neutron diffraction results of Soper and Phillips (dashed line).}
\label{fig:gootip4p}
\end{figure}

\begin{figure}[ht]
\include{gohtip4p}
\caption{Oxygen-hydrogen radial distribution function for the
TIP4P-FQ (solid line) and TIP4P (dotted line) potentials, comparing
to the neutron diffraction results of Soper and Phillips (dashed line).}
\label{fig:gohtip4p}
\end{figure}

\begin{figure}[ht]
\include{ghhtip4p}
\caption{Hydrogen-hydrogen radial distribution function for the
TIP4P-FQ (solid line) and TIP4P (dotted line) potentials, comparing
to the neutron diffraction results of Soper and Phillips (dashed line).}
\label{fig:ghhtip4p}
\end{figure}

\begin{figure}[ht]
\include{goospc}
\caption{Oxygen-oxygen radial distribution function for the
SPC-FQ (solid line) and SPC (dotted line) potentials, comparing
to the neutron diffraction results of Soper and Phillips (dashed line).}
\label{fig:goospc}
\end{figure}

\begin{figure}[ht]
\include{gohspc}
\caption{Oxygen-hydrogen radial distribution function for the
SPC-FQ (solid line) and SPC (dotted line) potentials, comparing
to the neutron diffraction results of Soper and Phillips (dashed line).}
\label{fig:gohspc}
\end{figure}

\begin{figure}[ht]
\include{ghhspc}
\caption{Hydrogen-hydrogen radial distribution function for the
SPC-FQ (solid line) and SPC (dotted line) potentials, comparing
to the neutron diffraction results of Soper and Phillips (dashed line).}
\label{fig:ghhspc}
\end{figure}

\begin{figure}[ht]
\include{pmu}
\caption{Distribution of the absolute value of the dipole moments
of the TIP4P-FQ (solid line) and SPC-FQ (dashed line) liquids.}
\label{fig:pmu}
\end{figure}

\clearpage
\newpage

\begin{figure}
\include{mut}
\caption{Time autocorrelation function for the system dipole for
the SPC-FQ (top line) and TIP4P-FQ (bottom line) models
and the exponential long-time approximation (dash lines),
shown on semilog axes.}
\label{fig:mut}
\end{figure}

\begin{figure}
\include{epst}
\caption{Real (top) and imaginary (bottom) parts of the frequency dependent
dielectric constant for the TIP4P-FQ model (solid lines),
compared to experiment (dotted lines).}
\label{fig:epst}
\end{figure}

\begin{figure}
\include{epss}
\caption{Real (top) and imaginary (bottom) parts of the frequency dependent
dielectric constant for the SPC-FQ model (solid lines),
compared to experiment (dotted lines).}
\label{fig:epss}
\end{figure}

\begin{figure}
\include{c2y}
\caption{Rotational correlation function for TIP4P-FQ (solid line) and
SPC-FQ (dotted line).}
\label{fig:c2y}
\end{figure}

\end{document}